\documentclass[12pt]{article}

\title{Gluelumps and Confinement in QCD}
\author{ Yu.A.Simonov\\
State Research
Center\\Institute of Theoretical and Experimental Physics, \\
Moscow, 117218 Russia}
\usepackage[dvips]{graphicx}
\newcommand{\beq}{\begin{eqnarray}}
 \newcommand{\eeq}{\end{eqnarray}}
\newcommand{\be}{\begin{equation}}
 \newcommand{\ee}{\end{equation}}

\def\ga{\mathrel{\mathpalette\fun >}}
\def\fun#1#2{\lower3.6pt\vbox{\baselineskip0pt\lineskip.9pt
\ialign{$\mathsurround=0pt#1\hfil ##\hfil$\crcr#2\crcr\sim\crcr}}}

\newcommand{{\SD}}{\rm SD}

\newcommand{\vesig}{\mbox{\boldmath${\rm \sigma}$}}

\newcommand{\veS}{\mbox{\boldmath${\rm S}$}}

\newcommand{\veB}{\mbox{\boldmath${\rm B}$}}

\newcommand{\veE}{\mbox{\boldmath${\rm E}$}}

\newcommand{\llan}{\langle\langle}
\newcommand{\rran}{\rangle\rangle}
\newcommand{\lan}{\langle}
\newcommand{\ran}{\rangle}

\begin{document}
\maketitle
\begin{abstract}Confinement   is explained via field correlators, and the
latter are calculated via gluelumps. Behavior of gluelump Green's function at
small and large distances yields gluonic condensate and vacuum correlation
length respectively and allows to check the consistency of the whole picture.

\end{abstract}

\section{Introduction}
 Field theory based on Abelian and nonabelian gauge fields is now the
 foundation of physics of  strong and electroweak interaction. Here  the gauge
 covariance  in  general and gauge invariance of physical amplitudes is the
 corner stone   and the most general principle, which allows to fix the form
 of  equations and resulting amplitudes even before any dynamical input. In the
 case of QCD this fact is even more important, as was stressed in \cite{1,2}.
 Below we show, that confinement makes gauge invariance and all relations,
 connected to it, the most important property sine qua non, and in turn, gauge
 invariance principle allows to build theory of confinement in good agreement
 with all existing data.

 It will be shown below, that gauge invariance requires each particle travelling
 from point $x$ to point $y$ to develop its own gauge factor $\Phi_C(x,y)$ (the
 Schwinger line, or parallel transporter) pertaining to the path $C(x,y)$. For
 neutral (color-singlet) systems the set of paths $\{C_i (x,y)\}$ forms a
 closed loop, a simple Wilson loop $W(C)$ for two opposite charges, or more
 complicated closed loop with $N_c$ branches for the $N_c$ charges in the group
 $SU(N_c)$.

 As will be shown (see also \cite{3} and \cite{4} for the review) all dynamics
 in the neutral system is given by the vacuum-averaged Wilson loop $\lan
 W(C)\ran$, and since all physical amplitudes  correspond to the gauge
 invariant and hence neutral systems, therefore $\lan W(C)\ran$ contains all
 the clues to confinement and in general to the dynamics of strong interaction.

Here one can  distinguish confining theories, like QCD, and nonconfining ones,
like electroweak (EW)  theory. In the  first case gauge invariance and
confinement make it impossible to consider only one part of color neutral
system, e.g. propagator of quark or gluon. Formally, gauge invariant vacuum
average of this colored object is zero, physically, the confining string, which
appears in $\lan W(C)\ran$ and connects this object to its neutralizing
partner, strongly governs its motion and cannot be disregarded. Therefore all
models, based on vacuum averages of gluon or quark  propagators in any gauge,
are irrelevant to the nature.

In nonconfining theories the situation is different. One can compose a system
of an electron here and proton on the moon and consider the motion of electron
in the gauge invariant language of $\lan W(C)\ran$, but the latter factorizes
into a product of propagators, however some initial and final conditions of
electron should be fixed to avoid divergencies.

A similar situation occurs in the deconfined phase of QCD \cite{5} (see
\cite{6} for a review), where (apart from magnetic confinement) $\lan W(C)\ran
$ also factorizes and dynamics is given by Wilson lines rather than Wilson
loops.

Thus the basic problem is first the analysis of $\lan W(C)\ran$ and, second,
self-consistent calculation of $\lan W(C)\ran$ or its ingredients. The first
part   is done using the   gauge invariant cumulant expansion of $\lan
W(C)\ran$ in terms of field correlators. As was shown in \cite{3,4,7}, for QCD
the lowest term (quadratic) $\lan F(x)F(y)\ran$ saturates  $\lan W(C)\ran$ with
few percent accuracy, which is supported by the Casimir scaling measurements on
the lattice both in the confined \cite{8} and deconfined phase \cite{9}. The
same Casimir scaling helps to put a strong upper limit on the presence of
topological charges or adjoint fluxes \cite{7} as a possible source of
confinement\footnote{ Note, that magnetic monopoles in QCD contribute  largely
 to higher cumulants and hence violate Casimir scaling (see \cite{9*}) while in the  weakly-coupled dilute $3d$  $SU(N)$
 Georgi-Glashow model the Casimir scaling holds to a good accuracy \cite{ad}, implying that higher cumulants are suppressed at low density.}.
 At the same time, Casimir scaling is well supported by the Gaussian quadratic
term $\lan F(x) F(y)\ran$ and the latter yields a beautiful picture of linear
confinement in good agreement with all lattice measurements and hadron physics.

Here comes the second part of problem: which field configurations are behind
the Gaussian term, and can one calculate these correlators selfconsistently, so
that all QCD can be derived from the one constant, fixing the scale of our
world. We show below, following \cite{10}, that this is in principle possible,
and one  can reexpress some physical observables, like string tension $\sigma$,
through others like $\Lambda_{QCD}$ (however still with low accuracy). More
importantly, one can establish the objects, which are behind confinement and
characterize full vacuum structure: those are known as  gluelumps\footnote{The
name and the first calculations in $SU(2)$ lattice theory are due to I.H.Jorysz
and C.Michael \cite{10*}. The detailed  $SU(3)$ calculations have been done in
\cite{20}}, and consist of one or two gluons propagating in the field of
adjoint static source. Mass of these objects was calculated analytically before
\cite{11} and its inverse gives the correlation length of the vacuum, while the
short distance behavior yields gluonic condensate, and finally, the integral of
gluelump Green's function yields string tension. Thus confinement is connected
to gluelumps, and all other vacuum properties can be calculated via gluelumps.
Moreover, the small correlation length $\lambda$ (due to large gluelump mass
$M, \lambda\sim  1/M$) explains, why the Casimir scaling holds in  QCD -- the
cumulant expansion is actually  a series in powers of the dimensionless
parameter $\sigma \lambda^2 \cong 0.05$ , and the first (quadratic) term is
dominant. Below we give a short exposition of Wilson loop and cumulant
expansion in section 2, gluelumps are introduced in section 3, and exploited in
section 4 to display consistency of the whole picture. Summary and conclusions
are given in section 5.

\section{Wilson loop and field Correlators}

The Green's function (propagator ) of quark can be written in the
Fock-Feynman-Schwinger Representation (FFSR) (see  \cite{12}  for a review) as
the quantum mechanical path integral (we consider Euclidean space-time
throughout the paper) \be S(x,y) = (m+\hat D)^{-1} = (m-\hat D) \int^\infty_0
ds (Dz)_{xy} e^{-K}P_A\exp (ig \int^x_y A_\mu dz_\mu) p_\sigma (x, y,s),
\label{1}\ee where $K= ms +\frac14 \int^s_0 \left(\frac{dz_\mu}{d\tau}\right)^2
d\tau$ and $W_\sigma= P_A\exp (ig \int^x_y A_\mu dz_\mu)p_\sigma$ is the
Schwinger line (parallel transporter) with spin insertions, and \be
p_\sigma(x,y,s)= P_F\exp \left[g\int^s_0 \sigma_{\mu\nu} F_{\mu\nu}
(z(\tau))d\tau\right],\label{2}\ee

The  last  factor  contains spin operators \be \sigma_{\mu\nu} F_{\mu\nu} =
\left(\begin{array}{ll} \vesig \cdot \veB& \vesig\cdot \veE\\ \vesig\cdot \veE&
\vesig\cdot \veB\end{array}\right), \label{3}\ee and will be important for
calculation of gluonic condensate in section 4; it also yields all
spin-dependent interactions in hadrons, see \cite{13} for a review and
references.

In  an analogous way one can construct FFSR for a gluon in the external and
vacuum fields

\be
 G(x,y) =
\int^\infty_0 ds (Dz)_{xy}  e^{-K}  P_A e^{ig\int^x_y \hat A_\mu dz_\mu}
p_\Sigma\label{4}\ee where $\hat A_\mu$ implies  $A_\mu$ in the adjoint
representation  and the spin factor $p_\Sigma$ can be written either in the
form (\ref{3}) with the gluon spin operator $\veS$ instead of $\vesig$, or
simpler, as \be p_\Sigma =[P_F\exp (2ig\int^s_0 F_{\lambda\sigma} (z(\tau))
d\tau)]_{\mu\nu}.\label{5}\ee We omit in this section the spin parts of quark
and gluon Green's function, since they do not define the confinement picture at
this stage.

For a white hadron, consisting of $q\bar q$ or two gluons one can write the
total Green's function as \be G_{q\bar q, gg} (\bar C_{xy})= \int
d\Gamma_{q\bar q, gg} (\bar C) W(\bar C_{xy})\label{6}\ee and the Wilson loop
operator for the closed loop $\bar C_{xy}$ is \be W(\bar C) = \lan P_A\exp (ig
\int_{\bar C} A_\mu dz_\mu)\ran,\label{7}\ee where the brackets imply vacuum
averaging. The use of the nonabelian Stokes theorem and cluster (cumulant)
expansion \cite{14} yields \be W(\bar C) = tr_c\exp \left\{\sum^\infty_{n=1}
\frac{(ig)^n}{n!} \int ds (1) ~~ \int ds (n) \llan F(1)... F(n)\rran\right\},
\label{8}\ee where $ds(k) = ds_{\mu_k\nu_k}$ is the surface element, and \be
F(k) \equiv F_{\mu_k\nu_k} (u_k, x_0) = \Phi (x_0, u_k) F_{\mu_k\nu_k} (u_k)
\Phi (u_k,x_0)\label{9}\ee while $\Phi(x,y)$ is the parallel transporter. Field
correlators $(FC) \llan F(1)... F(n)\rran$ depend in general on the chosen
point $x_0$, which is convenient to choose somewhere on the minimal surface
bounded by $\bar C$. The great simplification for QCD is that FC decrease so
fast with distance, as we shall see later, that $\Phi(u_k, x_0) \Phi(x_0, u_j)$
in (\ref{8},\ref{9}) can be replaced by straight line  transporter $\Phi(u_k,
u_j)$, and moreover, the whole series is fast converging as
$(\sigma\lambda^2)^n$, where $\lambda$ is the correlation length, $\lan F(x)
\Phi F (y) \Phi\ran \sim \exp \left( -\frac{|x-y|}{\lambda}\right).$ Therefore
one can keep only  the Gaussian correlator $\lan F\Phi F\Phi\ran$, and
moreover, since $\lambda$ is much smaller than  all other characteristic
quantities, e.g. hadron sizes, inverse masses of lowest hadrons etc., one can
say that the QCD vacuum configurations correspond to the Gaussian white noise.

As a result one can use only the lowest correlator

\be D_{\mu\nu, \lambda\sigma} \equiv \lan tr \frac{g^2}{N_c} F_{\mu\nu} (x)
\Phi(x,y) F_{\lambda\sigma} (y) \Phi(y, x)\ran,\label{10}\ee which can be
expressed in terms of two scalar functions $D(z)$ and $D_1(z)$
$$ D_{\mu\nu,\lambda\sigma} (z) = D(z) (\delta_{\mu\lambda
} \delta_{\nu\sigma} -\delta_{\mu\sigma} \delta_{\nu\lambda} ) +\frac12 \left[
\frac{\partial}{\partial z_\mu} ( z_\lambda \delta_{\nu\sigma} -z_\sigma
\delta_{\nu\lambda})+\right.$$ \be\left. + \frac{\partial}{\partial z_\nu} (
z_\sigma \delta_{\mu\lambda}
-z_\lambda\delta_{\mu\sigma})\right]D_1(z).\label{11}\ee

Note, that (\ref{10}), (\ref{11}) is written for fundamental charges, for
charge of irrepr. $j$ one should multiply by the  ratio $\frac{C(j)}{C({\rm
fund})}$, where $C(j)$ is the quadratic Casimir operator.

 Now all
the spin-independent dynamical properties can be calculated from static
potentials, the confining Lorentz scalar $V(r)$,

\be V(r) = 2\frac{C(j)}{C({\rm fund})}\int^r_0 (r-\lambda) d\lambda
\int^\infty_0 d\nu D(\sqrt{\lambda^2+\nu^2})\label{12}\ee and the Lorentz
vector $V_1(r)$ \be V_1(r) = \int^r_0 \lambda d\lambda \int^\infty_0 d\nu
D_1(\sqrt{\lambda^2+\nu^2})\label{13}\ee and the string tension (and
confinement) is due to $D(z)$: \be \sigma=\frac12\frac{C(j)}{C({\rm fund})}
\int\int d^2z D(z).\label{14}\ee Spin-dependent potentials can be found as well
for quarks of any mass or gluons without $1/M_Q$ expansion, keeping in
(\ref{8}) also spin terms, see \cite{13} for a review. The interaction
(\ref{12}), (\ref{13}) was used in the framework of the relativistic string
Hamiltonian \cite{15} and spectra of mesons \cite{16}, glueballs, \cite{17},
hybrids \cite{18} have been computed in excellent agreement with lattice and
experiment with minimal, input (current quark masses, $\sigma$ and $\alpha_s$
without any additional parameters) see \cite{19} for reviews.

At this point one could wonder how to check  the notion of Gaussian vacuum? The
answer is very simple: Gaussian correlators $D$ and $D_1$ are proportional to
$(gF)^2$, i.e. to the charge squared, and hence for higher color
representations $j$ they are proportional to the Casimir operator $C_2(j),$
known for the group $SU(N_c)$, $C_2({\rm fund}) =\frac{N^2_c-1}{2N_c}, ~~
C_2(adj) =N_c$ etc..

This constitutes the Casimir scaling, and the numerical  check  was done  on
the lattice \cite{8,9} and compared  with FC in \cite{7}. results of the
lattice measurements in the confined and deconfined phases  of QCD are shown in
Fig. 1 (taken from \cite{8}) and  Fig. 2 respectively. One can see an exellent
agreement of $V_{Q\bar Q} (j) = (- \frac{1}{T} \ln W_j(\bar C))$ computed on
the lattice for different $j=3,6,8,...$ with the Casimir scaled potential
$\frac{C_2(j)}{C_2(3)} V_{Q\bar Q} (3)$. (Note in Fig.1 $V$ and $r$ are units
of $r_0^{-1}$ and $r_0$ respectively, $r_0\approx 0.5$ fm.)

\begin{figure}[!h]\begin{center}
\includegraphics[width=7.5cm]{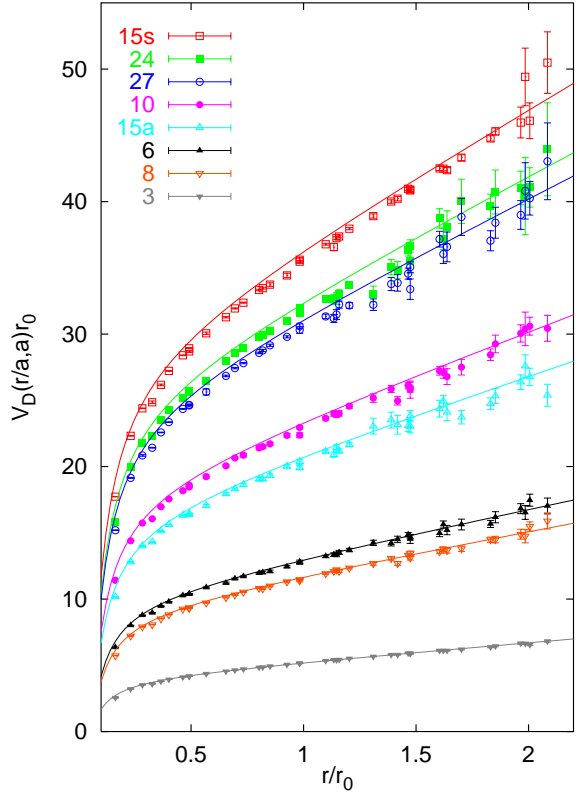}
\caption{(from [8]). The potentials $V_D$ in units of $1/r_0$  for all measured
representations $D=3,8,6,...$, obtained at $\beta=6.2$$(r_0=0.5$ fm,
$a=r_0/6.1)$. The line for a representation $D$ is  the  fundamental potential,
multiplied by the ratio $\frac{C_2(D)}{C_2({\rm fund})}.$}\end{center}
\end{figure}

\begin{figure}[t]\begin{center}
\includegraphics[width=10cm]{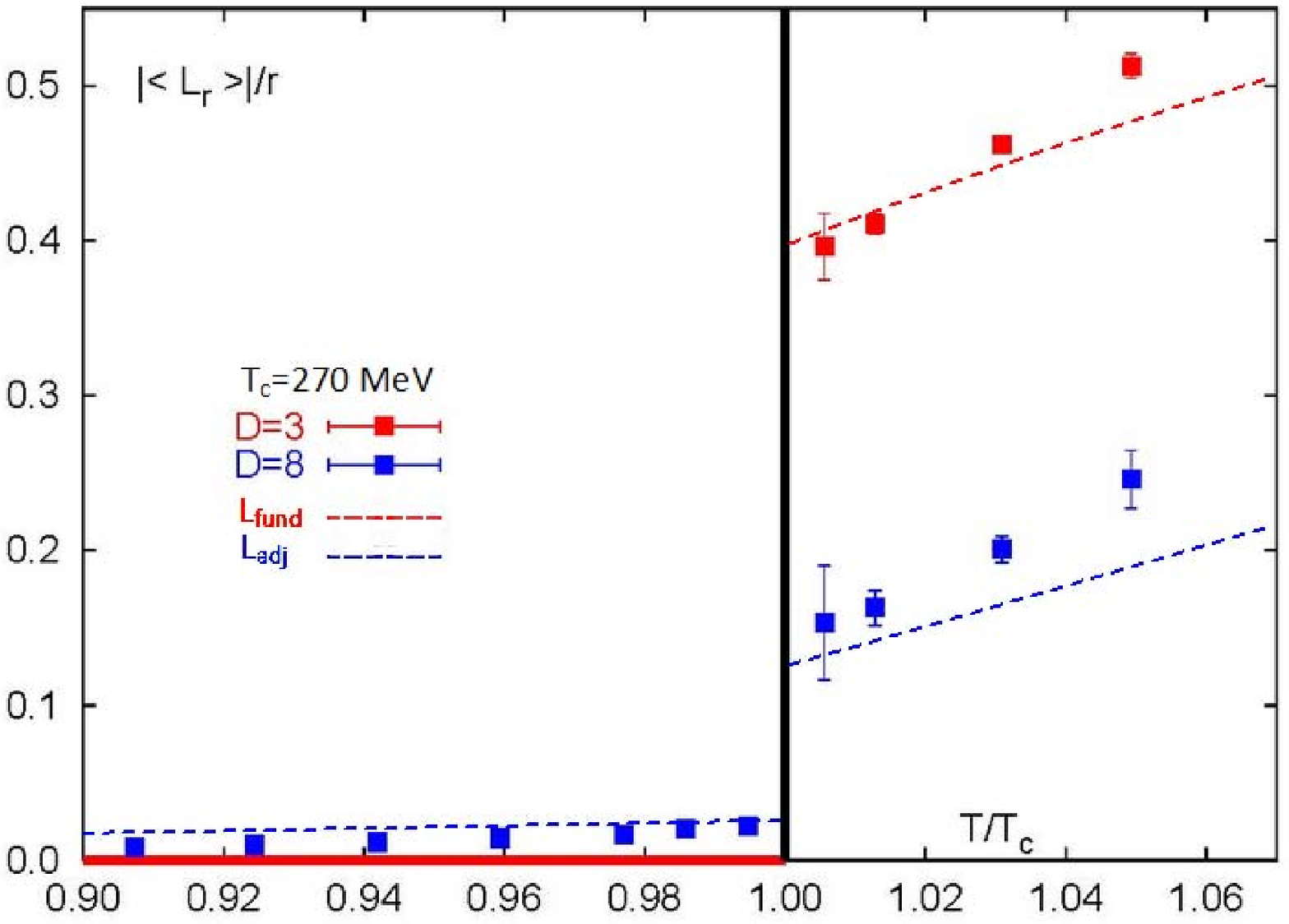}
\caption{Shown on the figure are curves of $L_{\rm adj}$ (lower curve for all
$T/T_c$) and $L_{\rm fund}$ (upper curve) compared to the lattice measured from
\cite{9}. In the $T<T_c$ region the $M(\bar{\alpha}_s=0.195)=0.982$ GeV
gluelump mass was used. In the deconfinement region the fit  was used with
$T_c=270$ MeV for $L_{\rm fund}$ and the Casimir scaled value for $L_{\rm
adj}$.}\end{center}
\end{figure}

In the deconfined phase only $V_1(r,T)$ is nonvanishing (apart from
colormagnetic correlators, see \cite{6} ) and it enters the Polyakov lines,
$L_j(T) =\exp \left(-\frac{C_2(j)}{C_2(3)}\frac{V_1(\infty,T)}{2T}\right)$. One
can see in Fig.2 the lattice measured  values\\ $ L_{\rm fund} (T), L_{\rm
adj}(T)$ as functions of temperature (points) vs our analytic calculations with
Casimir factors.

Agreement is impressive. (Note, that at $T<T_c$ in confinement phase $L_{\rm
fund} \equiv 0$, formally since one should replace $V_1\to V_1(\infty, T) +
V(\infty, T)$ and the latter is infinite).


\section{Gluelumps and Field Correlators}

To express FC through gluon propagators, one should consider gluons in
$D_{\mu\nu,\lambda\sigma}$, i.e.gluons emitted in $F_{\mu\nu} (x)$ and absorbed
in $F_{\lambda\sigma}(y)$ and moving in the background of   vacuum fields
(consisting of gluon fields again and with the source in $\Phi(x,y)$). This is
the basic idea of mean field approach. It is important, that we can express the
interaction of propagating gluons with background field again in terms of FC,
and thus one obtains a closed scheme, consistency of it will be discusses in
the next section. Expanding $F_{\mu\nu}$ into Abelian and  nonabelian parts,
$F_{\mu\nu}=(\partial_\mu A_\nu - \partial_\nu A_\mu) -ig [A_\mu,A_\nu]$ one
can write

\be D_{\mu\nu,\lambda\sigma} (x,y) = D^{(0)}_{\mu\nu,\lambda\sigma} +
D^{(1)}_{\mu\nu,\lambda\sigma} +D^{(2)}_{\mu\nu,\lambda\sigma}\label{15}\ee
where the superscript 0,1,2 refers to the power of $g$, coming from the term
$ig [A_\mu, A_\nu]$.

 Here $D^{(0)}_{\mu\nu,\lambda\sigma}$ is connected to the one-gluon gluelump
 Greens' function $G^{(1g)}_{\mu\nu}$,

\be D_{\mu\nu,\lambda\sigma}^{(0)} (x,y) =\frac{g^2}{2N^2_c}\left\{
\frac{\partial}{\partial x_\mu}\frac{\partial}{\partial y_\lambda}
G^{(1g)}_{\nu\sigma}(x,y) + {\rm perm}\right\} +
\Delta^{(0)}_{\mu\nu,\lambda\sigma},\label{16}\ee where the $1g$ gluelump
Green's function is \be G^{(1g)}_{\mu\nu} (x,y) = \lan {\rm Tr}_a A_\mu(x) \hat
\Phi (x,y) A_\nu (y)\ran.\label{17}\ee and $T_{r_a}$ implies summation over
adjoint indices.

$D_{\mu\nu,\lambda\sigma}^{(2)}$ is of basic importance, since  it ensures
confinement via $D(z)$ and is expressed via two-gluon gluelump Green's function
$G^{(2g)}(z)$

\be D^{(2)}_{\mu\nu,\lambda\sigma} (x,y) =(
\delta_{\mu\lambda}\delta_{\nu\sigma} - \delta_{\mu\sigma} \delta_{\nu\lambda})
D(x-y);~~ D(z) =\frac{g^4(N_c^2-1)}{2} G^{(2g)}(z),\label{18}\ee where both
one- and two-gluon gluelump Green's functions can be written in terms of path
integrals \cite{12} and finally expressed via eigenvalues and eigenfunctions of
relativistic string Hamiltonian \cite{15}. For gluelumps this Hamiltonian was
studied in \cite{11} and results for lowest gluelump masses in comparison with
lattice data \cite{20} are (expressed via fundamental string tension  $\sigma_f
=0.18$ GeV$^2$).

 $$M^{(1g)} = 1.49~{\rm GeV} , ~M^{(1g)} ({\rm lat}) < 1.7~{\rm GeV},~$$\be M^{(2g)} =
2.61~{\rm GeV}, ~ M^{(2g)} ({\rm lat}) = 2.7~{\rm GeV} ,\label{19}\ee where
$M^{(ig)}$ is the mass of gluelump with $i=1,2$ gluons. The same Hamiltonian
technic allows to find the asymptotics of Green's functions \cite{10}

\be D_1(z) = \frac{2C_2(f) \alpha_s M_0^{(1g)} \sigma_{adj}}{|z|}
e^{-M^{(1)}|z|},~~ |z| M^{(1g)} \gg1.\label{20}\ee \be D(z)
=\frac{g^4(N^2_c-1)}{2} 0.1 \sigma^2_f e^{-M^{(2g)} |z|},~~ M^{(2g)}
|z|\gg1\label{21}\ee Here $\alpha_s$ is the value of the strong coupling
constant at large distance.

Correspondingly, one obtains the correlation length of $D(z)$ and $D_1(z)$ \be
\lambda=\frac{1}{M^{(2g)}}\approx 0.08~{\rm fm},
\lambda_1=\frac{1}{M^{(1g)}}\cong 0.15~{\rm fm}.\label{22}\ee

This should be compared with $\lambda,\lambda_1$ calculated on the lattice
previously \cite{22}, $\lambda\approx \lambda_1\approx 0.2$ fm, while a recent
measurement \cite{23}, analyzed in \cite{13} yields $\lambda\approx 0.1$ fm,
and the numerical HP$^1$ formalism \cite{25} yields $\lambda\cong 0.13$ fm,
$\lambda_1\cong 0.145$ fm.

Finally, a remarkable agreement with the values (\ref{22}) has been found in
Ref. \cite{plb}, where the full calculation of the path integrals representing
$G_{\mu\nu}^{(1g)}$ and $G^{(2g)}$ has been performed.

Thus all data support the conclusion, that the correlation length is very small
$\lambda, \lambda_1 \sim 0.1$ fm and the QCD vacuum is therefore highly
stochastic. Moreover, we have calculated FC in terms of nonperturbative
parameters, like $\sigma,$ which can be calculated, as in (\ref{14}) via FC
themselves. This calls for the consistency check, to be done in the next
section.

\section{Consistency of the gluelump mechanism}

We start with small distances  and with $D_1(z)$.

The $1g$ gluelump Green's function can be written in FFSR \cite{12}

\be G^{(1g)}_{\mu\nu} (x,y) = {\rm Tr}_a \int^\infty_0 ds (Dz)_{xy} \exp(-K)
\lan W^F_{\mu\nu} (C_{xy})\ran,\label{23}\ee where the spin-dependent
Wilson-loop with insertion of operators $F_{\mu\nu}$ is  \be W^F_{\mu\nu}
(C_{xy})= PP_F \left\{ \exp (ig \int A_\lambda dz_\lambda) \exp
F\right\}_{\mu\nu}\label{24}\ee where $\exp F$ is as in (\ref{5}).

With zero background $(A_\mu\to 0)$ this is easily calculated and yields the
lowest order perturbative result, $D_1(z) =\frac{16\alpha_s}{3\pi z^4}$, then
from (\ref{13}) $V_1(r) =- \frac{4\alpha_s}{3r}$.

The first nonperturbative (np) contribution comes, when one accounts for the
small Wilson loop behaviour (second paper in \cite{3}) \be
W_{\mu\nu}^F\cong\exp \left( -\frac{\pi}{8} G_2 S^2\right), ~~ G_2
\equiv\frac{\alpha_s}{\pi} \lan (F^a_{\mu\nu} (0))^2\ran,~~ S~{\rm
area}.\label{25}\ee As was shown in \cite{10}, insertion of (\ref{25}) in
(\ref{23}) yields finally \be D_1(z) = \frac{16\alpha_s+O(\alpha^2_s\ln
z)}{3\pi z^4} + \frac{\pi^2}{6N_c} G_2.\label{26}\ee Note two important facts;
1) perturbative  and np contributions in (\ref{26}) enter additively in this
lowest order, this supports the original idea  in \cite{26}; 2) the sign of np
part is positive and this contributes to the np shift down of the vacuum energy
density.

We turn now to the function $D(z)$, connected via (\ref{18}) to the $2g$
gluelump Green's function. Here the small distance region is more complicated.
First of all, the same small-area term (\ref{25}) yields for $D(z)$ a {\bf
negative} result!

$$ D(z) = \Delta_1 D(z) + \Delta_2 D(z),$$
 \be \Delta_1D(z) =-\frac{g^4N_cG_2}{4\pi^2}\label{27}\ee However the same
spin-dependent Wilson loop (now with two gluons and one parallel transporter
(Schwinger line)) in (\ref{24}) contains paramagnetic contribution $\lan
W^F(1)W^F(2)\ran \sim \exp (-c\lan FF\ran)$, which has  {\bf positive} sign
(see Fig. 3).

\begin{figure}[h!]
\begin{center}
 \includegraphics[width=50mm,keepaspectratio=true]{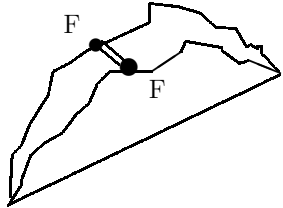}
 \caption{Interaction of two gluon trajectories, depicted by broken lines, via
 correlator of operators $F_{\mu\nu}$.}
\end{center}
\end{figure}

\be \Delta_2 D (z) =2N_c^2 g^4 h(z),~~h(z) \approx \frac{D(\lambda_0)}{64\pi^4}
\ln^2 \left( \frac{\lambda_0\sqrt{e}}{z}\right)\label{28}\ee

 Here   $\lambda_0
$ is the point, which is above the asymptotic free region of
$\alpha_s(\lambda(z))$, $\lambda_0\ga 0.24 $ fm (see Appendix of the last paper
in \cite{10}). Note an interesting feature of (\ref{28}): for vanishing $z$,
$g^2(z)\sim \frac{1}{\ln 1/z}$ and $\Delta_2 D(z)$ tends to a {\bf positive
constant}, so that the total contribution, \be D(z\to 0) =\Delta_1 D(z) +
\Delta_2D(z)\to D(0)=\frac{N^2_c}{2\pi^2}
D(\lambda_0)\left(\frac{2\pi}{\beta_0}\right)^2.\label{29}\ee

For $N_c=3$ this yields $D(0)\approx 0.15 D(\lambda_0)$, and implies the
humpbacked form of $D(z)$, see Fig.4, since it decreases exponentially at large
$z$, see Eq.(\ref{21}).

\begin{figure}[h!]
\begin{center}
 \includegraphics[width=45mm,keepaspectratio=true]{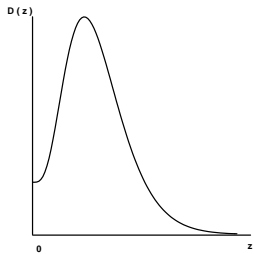}
 \caption{Qualitative behaviour of the  correlator $D(x)$ versus $x$, as
 predicted by (\ref{20}) and (\ref{27}),(\ref{28}) at large and small distances
 $x$ respectively.}
\end{center}
\end{figure}

Note, that this exponential behavior $D(z)=D_\sigma(0) e^{-|z|/\lambda}$, if
continued to zero, yields string tension from Eq.(\ref{14}) $\sigma\cong
\pi\lambda^2 D_\sigma(0)$, and $D_\sigma(0)=0.35$ GeV$^4$. Instead
$D(0)=\frac{\pi^2}{18} G_2$ and $G_2$ from \cite{26}(and this agrees with
transition temperature $T_c$, as shown in \cite{27}), $G_2\approx 0.01$
GeV$^4$, which yields $D(0)\sim 0.01$ GeV$^4$. Thus we understand
qualitatively, that this discrepancy between $D(0)$ and $D_\sigma(0)$ is due to
the AF behavior of (\ref{28}), illustrated by Fig.4.

Now let us discuss the selfconsistency at large distances. We start with the
asymptotic at $|z|\gg \lambda$, $D(z) =D_\sigma(0) \exp (-|z|/\lambda), ~~
\lambda=1/M_0^{(2)}$ with $D_\sigma(0)$ given in (\ref{21}), and assume, that
this form can be extended to $z=0$, at least in the integral
$\sigma=\frac12\int d^2zD(z)$, where small $z$ are less important.

As a result of this insertion of $D(z)$ into (\ref{14}), one obtains an
interesting relation \be 0.1\cdot 8\pi^2 \alpha^2_s (N_c^2-1)
\sigma^2_f=\frac{\sigma_f}{\pi\lambda^2}\label{30}\ee
 where in $\alpha_s=\alpha_s(\mu)$, and  the scale $\mu$ corresponds  to
 the  average momentum ($\approx$ inverse radius)  of the two-gluon gluelump. One can take $\mu_0\approx 1$ GeV, then from(\ref{30})
 one gets\footnote{Note the sign $\geq$ in (\ref{31}) due to  the fact, that
 the extension of exponential form of $D(z)$ to small $z$, overestimates the
 integral, cf. Eq(\ref{29})}
 \be
\alpha_s
 (\mu\cong 1~{\rm GeV})\geq  0.4;~~ \alpha_s(\mu)\cong \frac{4\pi}{\beta_0\ln
 \left(\frac{\mu^2+M^2_B}{\Lambda^2_{QCD}}\right)},\label{31}\ee
 where we have used the IR saturated form of $\alpha_s$, suggested earlier in
 \cite{28}, and derived for QCD with np background in \cite{29} and tested in
 hadronic physics in \cite{30} with the so-called background mass parameter  $M_B\cong 1$ GeV. Eq. (\ref{31}) is actually
 the ``dimensional transmutation relation'', since it connects $\sigma =0.18$
 GeV$^2$, as a basic scale parameter of QCD, yielding all masses and $\lambda$,
to $\alpha_s(\mu)$, i.e. $\Lambda_{\rm QCD}$. One gets from (\ref{31}) for
three flavors $n_f$
 \be \Lambda(n_f=3)\geq 0.29~{\rm GeV}.\label{32}\ee
 One should take into account, however, that we have used Eq.(\ref{31}) in the
 $x$-space, where $\lambda$ should be approximately 1.3 times larger, than
 $\Lambda_{\rm QCD} (n_f=3)\approx 0.27$ GeV.

 Thus this large distance consistency check is reasonably satisfied.

 We end this section with two remarks:

 1) Till now only  np contribution to $D(z)$ have been discussed. Naively the
 2g gluelump Green's function yields perturbative result for $D(z) \sim
 \frac{g^4}{z^4}$, however, as shown in \cite{31}, all perturbative
 contributions to $D(z)$ are exactly cancelled by those from higher
 correlators.

2) $D(z) $ and $D_1(z)$ have been obtained here in the leading approximation,
when gluelumps of minimal number of gluons contribute: 2 for $D(z)$ and 1 for
$D_1(z)$. In the higher orders of $O(\alpha_s)$ one has an expansion of the
type
$$ D(z) = D^{(2gl)} (z) + c_1 \alpha_s^3 D^{(1gl)} (z) + c_2
\alpha^3_s D^{(3gl)} (z) +...$$ \be D_1(z) =D^{(1gl)} (z) + c'_1 \alpha^3_s
D^{(2gl)} (z) +...\label{33}\ee Hence the asymptotic behavior for $D(z)$ will
contain exponent of $M_0^{(1)} |z|$ too, but with a small  preexponent
coefficient.

\section{Summary and discussion}

We have shown, that every physical amplitude can be expressed through the
corresponding Wilson loop, integrated as a path integral with the known weight.
Therefore all perturbative  and nonperturbative expressions for any amplitude
can be derived from the corresponding   expansions of the Wilson loop, e.g. the
double-logarithm asymptotics of Sudakov type also follows from this
representation \cite{32}. On the other hand, Wilson loop can be exactly given
by the cumulant series, containing all FC, and this is exact representation
both for perturbative and np contributions.

We have also demonstrated, that the np part of the Wilson loop is saturated by
the quadratic (Gaussian) FC, which is proved by Casimir scaling on the lattice
and understood as an expansion in powers of $(\sigma\lambda)^2\cong0.04 \ll 1.$

Thus all observables can be expressed via Gaussian FC with accuracy of few
percent. On the other  hand we can calculate FC as gluelump Green's functions,
and the latter are computed again in terms of FC (or simply in terms of
$\sigma,$ which is integral of FC). Thus one obtains the closed scheme of
calculation and approximations, which is subject to the selfconsistency check.
This analysis reveals an interesting picture, explaining why gluonic condensate
is relatively weak (as compared with hadron mass scale), and supporting
quantitatively the idea of positive $G_2$ and hence negative vacuum energy
density, yielding np stable vacuum. Finally, the idea of QCD as selfconsistent
theory defined by the only one mass scale, is realized in this FC approach and
the first connection between $\sigma$ and $\Lambda_{QCD}$, exemplifying
dimensional transmutation, is obtained in (\ref{31}).

The approach, discussed above, is universal and can be applied both to spectrum
and scattering, in particular to high-energy processes, where np contributions
are important. In all cases the problem reduces to the ground state or excited
string dynamics, pertinent to the Wilson loop manifold with changeable (and
integrated over) contours. Some first steps in this direction were done in
\cite{33,12}.

\section{Conclusion}

The modern picture of confinement in QCD is a result of tedious efforts of many
people during last 35 years, see \cite{34,35} for reviews. The basic source of
 confinement -- the  nonperturbative vacuum -- has passed many stages during
 this period. From constant chromomagnetic vacuum of Savvidy \cite{36}, and
 selfdual of Leutwyler \cite{37}, via spaghetti-vacuum of Nielsen-Olesen
 \cite{38}, using general idea of stochastic vacuum of Ambjorn,  Olesen,
 Peterson \cite{39} and finally to the Gaussian stochastic vacuum of the Field
 Correlator Method \cite{3,4}, discussed in this short review. Many people
 contributed to this field,  suggesting models  of  confinement, based on different
 mechanisms (see discussion in reviews \cite {34,35}), which we could not
 include here, and this study will certainly go on. However, at present the
 only approach, which explains all details of confinement in experiment and
 lattice calculation, is the Field Correlator Method (FCM). The Gaussian
 correlators $D$ and $D_1$ (and their colormagnetic counterparts  for nonzero temperatures) are sufficient to describe all data up
 to now. The fact, that it is possible to calculate both $D$ and $D_1$
 selfconsistently, with the only one scale input (in addition to current quark
 masses), gives a strong hope to construct the analytic nonperturbative QCD ab
 initio, as a valuable counterpart of lattice QCD.

 The author is grateful to D.Antonov for discussions and  valuable suggestions.
 The author acknowledges financial support of the RFFI grant 09-02-00629a.

\end{document}